\documentclass[ aps, showpacs, showkeys, nofootinbib, floatfix, superscriptaddress]{revtex4}

\usepackage{amsfonts}

\usepackage{amssymb}
\usepackage{amsmath}
\usepackage{graphicx}

\begin{document}

\title{Torsion and accelerating expansion of the universe in quadratic
gravitation}

 \author{Guoying Chee}
 \email{qgy8475@sina.com}
 \affiliation{College of physics and electronics, Liaoning Normal University,
Dalian,\\
116029, China,\\
Purple Mountain Observation, Academia Sinica, Nanjing, 210008, China}
\author{Yongxin Guo}
\affiliation{Physics Department, Liaoning University, Shenyang 110036, China}

\begin{abstract}
 Several exact cosmological solutions of a metric-affine theory of gravity
with two torsion functions are presented. These solutions give a essentially
different explanation from the one in most of previous works to the cause of
the accelerating cosmological expansion and the origin of the torsion of the
spacetime. These solutions can be divided into two classes. The solutions in
the first class define the critical points of a dynamical system
representing an asymptotically stable de Sitter spacetime. The solutions in
the second class have exact analytic expressions which have never been found
in the literature. The acceleration equation of the universe in general
relativity is only a special case of them. These solutions indicate that
even in vacuum the spacetime can be endowed with torsion, which means that
the torsion of the spacetime has an intrinsic nature and a geometric origin.
In these solutions the acceleration of the cosmological expansion is due to
either the scalar torsion or the pseudoscalar torsion function. Neither a
cosmological constant nor dark energy is needed. It is the torsion of the
spacetime that causes the accelerating expansion of the universe in vacuum.
All the effects of the inflation, the acceleration and the phase
transformation from deceleration to acceleration can be explained by these
solutions. Furthermore, the energy and pressure of the matter without spin
can produce the torsion of the spacetime and make the expansion of the
universe decelerate as well as accelerate.

\end{abstract}



\maketitle

\section{$\text{Introduction}$}

 In the last few years the realization that the universe is currently
undergoing an accelerated expansion phase and the quest for the nature of
dark energy has renewed interest in so-called modified gravity theories (for
a review see [1]). In these theories one modifies the laws of gravity so
that a late-time accelerated expansion is produced without recourse to a
dark energy component, a fact which renders these models very attractive.
The simplest family of modified gravity theories is obtained by replacing
the Ricci scalar $R$ in the usual Hilbert- Einstein Lagrangian with some
function $f(R)$ (for reviews, see , [2-5]).

There are actually three versions of $f(R)$ gravity: Metric $f(R)$ gravity,
Palatini $f(R)$ gravity, and metric-affine $f(R)$ gravity. In fact, these
are physically different theories rather than manifestations of the same
theory in different guises, as the different variational principles yield
inequivalent equations of motion (except when the action is the
Einstein-Hilbert and matter is minimally coupled to geometry). In metric $%
f(R)$ gravity, the action is varied with respect to the metric as usual (for
an introduction see [6]). Palatini $f(R)$ gravity comes about from the same
action if we decide to treat the connection as an independent quantity. The
connection, however, does not enter the matter action. Such a approach were
introduced and initially studied by Buchdahl [7] and has attracted a lot of
interest as possible infrared modifications of general relativity (for a
shorter review of metric and Palatini $f(R)$ gravity see [8]). It has
recently been generalized to $f(R)$ theories with non-symmetric connections,
i.e. theories that allow for torsion [9] and $f(R,R_{\mu \nu }R^{\mu \nu })$
theories [10]. In metric-affine $f(R)$ gravity the matter action is allowed
to depend also on the connection. In addition, the connection{\em \ }can
include both torsion and non-metricity [11].

It has been shown that even in the most general case of Palatini $f(R)$
gravity where both torsion and non-metricity are allowed, the connection can
still be algebraically eliminated in favor of the metric and the matter
fields [12]. Clearly, $f(R)$ actions do not carry enough dynamics to support
an independent connection which carries dynamical degrees of freedom.
However, this is not a generic property of generalized Palatini gravity.
{\em \ }The addition of the $R_{\mu \nu }R^{\mu \nu }$ term to the
Lagrangian radically changes the situation and excites new degrees of
freedom in the connection. The connection (or parts of it) becomes dynamical
and so, it cannot be eliminated algebraically. If the connection is torsion
free, the dynamical degrees of freedom reside\ in the symmetric part of the
connection [13].

In generic metric-affine theories the addition of the $R_{\mu \nu }R^{\mu
\nu }$ term to the Lagrangian makes the propagating degrees of freedom
reside in both the antisymmetric and symmetric parts of the connection. In
other words, the dynamical degrees of freedom can be both torsion and
non-metricity. In these theories torsion field plays a fundamental role: it
contributes, together with curvature degrees of freedom, to the dynamics.
Propagating torsion is the key feature of these theories [14, 15].

Torsion proves to be essential for total angular momentum conservation when
intrinsic spin angular momentum is relevant (for reviews on torsion, see
[16, 17, 18]). It has been argued that torsion must be present in a
fundamental theory of gravity [19, 20]. In the teleparallel gravity, for
example, torsion plays a central role (for a shorter review see [21]).
Recently, models based on modified teleparallel gravity, namely $f(T)$, were
presented. In these models the\ torsion proves to be the responsible of the
observed acceleration of the universe [22].

The rediscovery of the metric-affine (Palatini) formulation was mainly
driven by the interest in finding cosmological scenarios able to explain the
current observations. Using the respective dynamically equivalent
scalar-tensor representation of Palatini $f(R)$ gravity some cosmological
models with asymptotically de Sitter behavior have been presented [23]. It
was shown that adopting the metric-affine formulation together with an
action that includes a term{\em \ }inversely proportional to the scalar
curvature, such as the one in [24], can address the problem of the current
accelerated expansion equally well as when using the purely metric formalism
[25]. Additionally, it was found that $f(R)$ theories of gravity in the
metric-affine formulation do not suffer from the problems for the metric
formulation. On the other hand, although cosmology in the theories with the
Lagrangian including $R^2$ and $R_{\mu \nu }R^{\mu \nu }$ terms have been
studied in the purely metric formulation (for example see [26]) and the
Palatini formulation [27], the similar cosmological models in the
metric-affine formulation have not been discussed thoroughly in the
literature. Especially, the cosmological effect of torsion in metric-affine
theories of gravity has not been explored extensively. We have not known
whether the dynamical torsion could lead to a de Sitter solution and then be
used to explain the observed acceleration of the universe.{\em \ }An answer
will be given in this paper.

The metric-affine approach has been widely used in order to interpret
gravity as a gauge theory many times over the years (see, for example, [28]
for a study on $f(R)$ actions and [29] for a thorough review). In recent
years it has bee used in cosmology to interpret the accelerating expansion
of the universe [30, 31]. In this approach the structure of the
gravitational equations and physical consequences of cosmology, in
particular, the situation concerning the accelerating expansion depend
essentially on the form of the Lagrangian. The metric-affine gravity can be
divided into different sectors in dependence on the number of nonvanishing
components of the torsion tensor and the order of the differential
equations. One sector of the metric-affine gravity is so-called dynamical
scalar torsion sector considered in [30]. Starting from a Lagrangian
consisting of $R^2$ and the quadratic torsion terms a cosmological model has
been constructed. This model can contribute an oscillating aspect to the
expansion rate of the universe. A different model of acceleration with
torsion but without dark matter and dark energy has been presented in [31].
The Lagrangian of it is the most general form including the linear in the
scalar curvature term as well as 9 quadratic terms (6 invariants of the
curvature tensor and 3 invariants of the torsion tensor with indefinite
parameters). Its Lagrangian involves too many terms and indefinite
parameters, which make the field equations complicated and difficult to
solve and the role of each term obscure. In order to simplify the field
equations some restrictions on indefinite parameters have to be imposed.
Under these restrictions, especially, all the higher derivatives of the
scale factor are excluded from the cosmological equations. The question is
whether such a complicated Lagrangian is necessary. Can we use a simpler
Lagrangian to construct a model of cosmic acceleration? In fact all the
indefinite parameters in the Lagrangian in [31] have been combined into four
new ones, which implies that some terms are not necessary and the Lagrangian
can be simplified. In this paper we will show that a rather simpler
Lagrangian, $R+\alpha R^2+\beta R_{\mu \nu }R^{\mu \nu }+\gamma T{}^\mu
{}_{\nu \rho }T{}_\mu {}^{\nu \rho }$, is sufficient and necessary to
construct a model of cosmic acceleration. The terms $\beta R_{\mu \nu
}R^{\mu \nu }$ and $\gamma T{}^\mu {}_{\nu \rho }T{}_\mu {}^{\nu \rho }$
play different roles in the theory: the former determines the structure of
the field equations while the latter determines the behavior and the
stability of the solutions. The $\beta R_{\mu \nu }R^{\mu \nu }$ term leads
to different structure of the cosmological equations from the one in [30].
In addition to the simplicity the main advantage of this Lagrangian is to
permit exact or analytic solutions which have not been found in previous
works. For any physical theories, to find exact or analytic solutions is an
important topic. Next comes the physical interpretation of the solutions
thus obtained. Mathematically de Sitter spacetime as the maximally space is
undoubtedly important for any gravity theories. From the observational side,
recent studies illuminate that both the early universe (inflation) and the
late-time universe (cosmic acceleration) can be regarded as fluctuations on
a de Sitter background. So de Sitter solutions take a pivotal status in
gravitational theories, especially in modern cosmology.

We will follow the approach of [26, 30,31] rather than the one in [27] to
avoid getting involved in debate on the transformation from one frame to
another[32]. We choose the tetrad$ e_I^{\mu}$ and the spin
connection $\Gamma {}^{IJ}{}_\mu $ instead of the metric $g_{\mu \nu }$ and
the affine connection $\Gamma {}^\lambda {}_{\mu \nu }$ as the dynamical
variables following the gauge theory approach [30]. The descriptions in
terms of the variables ($e^I{}_\mu $, $\Gamma {}^{JK}{}_\nu $) and ($g_{\mu
\nu }$, $T^\lambda {}_{\rho \sigma }$) are equivalent in our approach (the
argument in detail see [16]). We will concentrate on the role of torsion
subject to the metricity. In this case only the torsion part
of the connection is independent of the metric (or tetrad).

Because the field equations can result of order higher than second and very
difficult to handle, the theory of dynamical systems provides a powerful
scheme for investigating the physical behavior of such theories [33] for a
wide class of cosmological models. The dynamical system approach has
acquired great importance in the investigation on various theories of
gravity. Some works have been done in the case of scalar fields in cosmology
and for scalar-tensor theories of gravity [34]. This approach has the
advantage of offering a relatively simple method to obtain exact solutions
(even if these only represent the asymptotic behavior) and to obtain a
(qualitative) description of the global dynamics of the models. Such results
are very difficult to obtain by other methods. The application of this
method has allowed new insights on higher order cosmological models and has
shown a deep connection between these theories and the cosmic acceleration
phenomenon. It makes possible not only to develop experimental test for
alternative gravity but also to allow a better understanding of the reasons
underlying the success of the theory. The dynamical systems approach has
been used to investigate universes in theories of gravity [26,30, 31]. In
contrast with [31] we allow the field equations to contain higher
derivatives. We will see that for the Lagrangian of the form $R+\alpha
R^2+\beta R_{\mu \nu }R^{\mu \nu }+\gamma T{}^\mu {}_{\nu \rho }T{}_\mu
{}^{\nu \rho }$ the field equations can be simplified and solved exactly for
some choices of $\alpha $, $\beta $ and $\gamma $. Some meaningful
consequences can be inferred from the solutions obtained. The accelerating
expansion of the universe can be explained without a cosmological constant
or dark energy. A vacuum spacetime can possess torsion which causes the
acceleration of the cosmological expansion. The conception of vacuum as
physical notion is changed essentially. Instead of it as passive receptacle
of physical objects and processes, the vacuum assumes a dynamical properties
as a gravitating object. The torsion of the spacetime can be produced by the
energy and pressure besides the spin of matter.

The paper is organized as follows. In section II the gravitational field
equations are derived following the approach of [29, 30, 31]. Using them to
the spatially flat Friedmann-Robertson-Walker metric a system of
cosmological equations is obtained in section III. Since the spin
orientation of particles in ordinary matter is random, the macroscopic
spacetime average of the spin vanishes. In this case, the solutions of the
cosmological equations are divided into two classes. Each of them is related
with only one torsion function, the scalar or the pseudoscalar torsion
function. They are obtained in section IV and V, separately, using different
methods. For the scalar torsion function the equations take the form of a
dynamical system, of which asymptotically stable critical points represent
the exact de Sitter solutions. For the pseudoscalar torsion function an
exact analytic solution of the cosmological equations is presented in
section V. In terms of this solution the acceleration and the phase
transformation from decelerating to accelerating expansion of the universe
can be explained. All of these solutions indicate that in vacuum the
spacetime possesses an intrinsic torsion which does not originate from the
spin of matter. It is the torsion that causes the acceleration of the
cosmological expansion in vacuum. The torsion of the spacetime can be
produced by the energy and pressure besides the spin of matter. In section
VI we obtain some exact analytic solutions of the cosmological equations in
the case $\gamma =0$. These solutions can only describe the inflation (in
the early epoch) or the decelerating expansion (in the later epoch) of the
universe. This means that the term $\gamma T{}^\mu {}_{\nu \rho }T{}_\mu
{}^{\nu \rho }$ is necessary to construct a model of cosmic acceleration.
The section VII is devoted to conclusions.

\section{$\text{Gravitational field equations}$}

We start from the action
\begin{equation}
S\left[ g_{\mu \nu },\Gamma ^\alpha {}_{\beta \gamma },\psi \right] =\frac %
\hbar {8\pi l^2}\int d^4x\sqrt{-g}\left[ \frac 12R+\alpha R^2+\beta R_{\mu
\nu }R^{\mu \nu }+\gamma T{}^\mu {}_{\nu \rho }T{}_\mu {}^{\nu \rho }\right]
+S_m\left[ g_{\mu \nu },\Gamma ^\alpha {}_{\beta \gamma },\psi \right] ,
\end{equation}
where $l=\sqrt{\hbar G/c^3}$ is the Planck length, $\alpha $, and $\beta $
are two parameters with the dimension of $l^2$, $\gamma $ is a parameter of
dimensionless, $\psi $ denotes matter fields. In contrast with [27], here
the connection is not symmetric, i.e. $\Gamma ^\alpha {}_{\beta \gamma }\neq
\Gamma ^\alpha {}_{\gamma \beta }$, and appears in the action of matter $S_m$%
. In other words, we are dealing with a metric-affine theory rather than a
Palatini one. By the same way used in [29, 30, 31], the variational
principle yields the field equations for the tetrad{\em \ } $e_I{}^\mu $ and
the spin connection $\Gamma {}^{IJ}{}_\mu $:

\begin{eqnarray}
&&e^{I\nu }R{}_{\nu \mu }-\frac 12e{}^I{}_\mu R  \nonumber \\
&=&E^I{}_\mu -\alpha \left( 4e^{I\nu }R{}_{\nu \mu }-e{}^I{}_\mu R\right)
R-\beta \left( 2e{}^{I\sigma }R{}^\rho {}_\sigma R{}_{\rho \mu }+2e^J{}_\rho
R{}^{\rho \sigma }{}R{}^I{}_{J\mu \sigma }-e{}^I{}_\mu R_{\rho \sigma
}R^{\rho \sigma }\right)  \nonumber \\
&&+\gamma \left( 4\partial _\nu \left( e{}^{I\lambda }T{}_{\mu \lambda
}{}^\nu \right) -4e{}^K{}_\tau e{}^{I\lambda }T{}_{\mu \lambda }{}^\nu
\partial _\nu e_K{}^\tau +e{}^I{}_\mu T{}^\lambda {}_{\rho \sigma
}T{}_\lambda {}^{\rho \sigma }-4e{}^{I\nu }T{}^\lambda {}_{\nu \tau
}T{}{}_{\lambda \mu }{}^\tau \right) ,
\end{eqnarray}
\begin{eqnarray}
&&e_{[I}{}^\nu e_{J]}{}^\mu e{}^K{}_\tau \partial _\nu e_K{}^\tau
+e_{[I}{}^\nu e_{J]}{}^\tau \Gamma ^\mu {}_{\nu \tau }+e_{[I}{}^\nu
e_{J]}{}^\mu \Gamma ^\lambda {}_{\lambda \nu }  \nonumber \\
&=&s_{IJ}{}^\mu -4\alpha \left( e_{[I}{}^\nu e_{J]}{}^\tau \Gamma ^\mu
{}_{\nu \tau }R+e_{[J}{}^\mu e_{I]}{}^\nu \left( \Gamma ^\lambda {}_{\lambda
\nu }R-\partial _\nu R\right) +e_{[I}{}^\nu e_{J]}{}^\mu Re{}^K{}_\tau
\partial _\nu e_K{}^\tau \right)  \nonumber \\
&&-4\beta e_J{}^\lambda \left( e_I{}^{[\mu }\partial _\nu R_\lambda {}^{\nu
]}+e_I{}^{[\nu }R_\lambda {}^{\mu ]}e{}^K{}_\tau \partial _\nu e_K{}^\tau
+e_I{}^\tau \Gamma ^{[\nu }{}_{\nu \tau }R_\lambda {}^{\mu ]}+e_I{}^{[\nu
}R_\tau {}^{\mu ]}\Gamma ^\tau {}_{\nu \lambda }\right)  \nonumber \\
&&-4\gamma e_{I\nu }e{}_J{}^\tau T{}^{\nu \mu }{}_\tau .
\end{eqnarray}
where $E^I{}_\mu $ and $s_{IJ}{}^\mu $ are energy- momentum and spin tensors
of the matter source, respectively. We use the Greek alphabet ($\mu $, $\nu $%
, $\rho $, $...=0,1,2,3$) to denote (holonomic) indices related to
spacetime, and the Latin alphabet ($I,J,K,...=0,1,2,3$) to denote algebraic
(anholonomic) indices, which are raised and lowered with the Minkowski
metric $\eta _{IJ}$ $=$ diag ($-1,+1,+1,+1$). If $\alpha =\beta =\gamma =0$,
these equations become the field equations of Einstein-Cartan-Sciama-Kibble
theory.. Especially, (20) becomes the Einstein equation. To understand these
equations, we will do a translation of (2, 3) into a certain effective
Riemannian form--transcribing from quantities expressed in terms of the
tetrad $e_I{}^\mu $ and spin connection $\Gamma {}^{IJ}{}_\mu $ into the
ones expressed in terms of the metric $g_{\mu \nu }$ and torsion $T^\lambda
{}_{\mu \nu }$ (or contortion $K^\lambda {}_{\mu \nu }$), as was done in
[30]. It should be noted [16] that the set ($e^I{}_\mu $, $\Gamma
{}^{JK}{}_\nu $) corresponds to the first order formalism, while the set ($%
g_{\mu \nu }$, $T^\lambda {}_{\rho \sigma }$) to the second order formalism.
The origin of this is that in the last case the non-torsional part of the
affine connection is a function of the metric, while, within the gauge
approach, the variables ($e^I{}_\mu $, $\Gamma {}^{JK}{}_\nu $) are mutually
independent completely. The descriptions in terms of the variables ($%
e^I{}_\mu $, $\Gamma {}^{JK}{}_\nu $) and ($g_{\mu \nu }$, $T^\lambda
{}_{\rho \sigma }$) are equivalent in our approach (the argument in detail
see [16]).

Subject to the metricity,{\em \ } the affine connection $\Gamma ^\lambda
{}_{\mu \nu }$ is related to the tetrad{\em \ } $e_I{}^\mu $ and the spin
connection $\Gamma {}^{IJ}{}_\mu $ by
\begin{eqnarray}
\Gamma ^\lambda {}_{\mu \nu } &=&e_I{}^\lambda \partial _\mu e^I{}_\nu
+e_J{}^\lambda e^I{}_\nu \Gamma {}^J{}_{I\mu }  \nonumber \\
&=&\left\{ _\mu {}^\lambda {}_\nu \right\} +K^\lambda {}_{\mu \nu },
\end{eqnarray}
where $\left\{ _\mu {}^\lambda {}_\nu \right\} $, $K^\lambda {}_{\mu \nu }$
are the Levi-Civita connection and the contortion, separately, with
\begin{eqnarray}
K^\lambda {}_{\mu \nu } &=&-\frac 12\left( T^\lambda {}_{\mu \nu }+T_{\mu
\nu }{}^\lambda +T_{\nu \mu }{}^\lambda \right) ,  \nonumber \\
T^\lambda {}_{\mu \nu } &=&e_I{}^\rho T^I{}_{\mu \nu }=\Gamma ^\lambda
{}_{\mu \nu }-\Gamma ^\lambda {}_{\nu \mu },  \nonumber \\
T{}^I{}_{\mu \nu } &=&\partial _\mu e{}^I{}_\nu -\partial _\nu e{}^I{}_\mu
+\Gamma {}^I{}_{J\mu }e{}^J{}_\nu -\Gamma {}^I{}_{J\nu }e{}^J{}_\mu .
\end{eqnarray}
Accordingly the curvature $R^\rho {}_{\sigma \mu \nu }$ can be represented
as
\begin{eqnarray}
R^\rho {}_{\sigma \mu \nu } &=&e_I{}^\rho e^J{}_\sigma R^I{}_{J\mu \nu
}=\partial _\mu \Gamma ^\rho {}_{\sigma \nu }-\partial _\nu \Gamma ^\rho
{}_{\sigma \mu }+\Gamma ^\rho {}_{\lambda \mu }\Gamma ^\lambda {}_{\sigma
\nu }-\Gamma ^\rho {}_{\lambda \nu }\Gamma ^\lambda {}_{\sigma \mu }
\nonumber \\
&=&R_{\left\{ {}\right\} }^\rho {}_{\sigma \mu \nu }+\partial _\mu K^\rho
{}_{\sigma \nu }-\partial _\nu K^\rho {}_{\sigma \mu }+K^\rho {}_{\lambda
\mu }K^\lambda {}_{\sigma \nu }-K^\rho {}_{\lambda \nu }K^\lambda {}_{\sigma
\mu }  \nonumber \\
&&+\left\{ _\lambda {}^\rho {}_\mu \right\} K^\lambda {}_{\sigma \nu
}-\left\{ _\lambda {}^\rho {}_\nu \right\} K^\lambda {}_{\sigma \mu
}+\left\{ _\sigma {}^\lambda {}_\nu \right\} K^\rho {}_{\lambda \mu
}-\left\{ _\sigma {}^\lambda {}_\mu \right\} K^\rho {}_{\lambda \nu },
\end{eqnarray}
where $R_{\left\{ {}\right\} }^\rho {}_{\sigma \mu \nu }=\partial _\mu
\left\{ _\sigma {}^\rho {}_\nu \right\} -\partial _\nu \left\{ _\sigma
{}^\rho {}_\mu \right\} +\left\{ _\lambda {}^\rho {}_\mu \right\} \left\{
_\sigma {}^\lambda {}_\nu \right\} -\left\{ _\lambda {}^\rho {}_\nu \right\}
\left\{ _\sigma {}^\lambda {}_\mu \right\} $ is the Riemann curvature of the
Levi-Civita connection. In view of this, we can identify the actual degrees
of freedom of the theory with the (independent) components of the metric $%
g_{\mu \nu }$ and the tensor $K^\lambda {}_{\mu \nu }$.

\section{$\text{Cosmological equations}$}

For the spatially flat Friedmann-Robertson-Walker metric
\begin{equation}
g_{\mu \nu }=\text{diag}\left( -1,a\left( t\right) ^2,a\left( t\right)
^2,a\left( t\right) ^2\right) ,
\end{equation}
the non-vanishing components of the Levi-Civita connection are
\begin{eqnarray}
\left\{ _0{}^0{}_0\right\} &=&0,\left\{ _0{}^0{}_i\right\} =\left\{
_i{}^0{}_0\right\} =0,\left\{ _i{}^0{}_j\right\} =a\stackrel{\cdot }{a}%
\delta _{ij},  \nonumber \\
\left\{ _0{}^i{}_0\right\} &=&0,\left\{ _j{}^i{}_0\right\} =\left\{
_0{}^i{}_j\right\} =\frac{\stackrel{\cdot }{a}}a\delta _j^i,\left\{
_j{}^i{}_k\right\} =0,i,j,k,...=1,2,3.
\end{eqnarray}
The non-vanishing torsion components with holonomic indices are given by two
functions, the scalar torsion $h$ and the pseudoscalar torsion $f$ [35]:
\begin{eqnarray}
T_{110} &=&T_{220}=T_{330}=a^2h,  \nonumber \\
T_{123} &=&T_{231}=T_{312}=2a^3f,
\end{eqnarray}
and then the contortion components are
\begin{eqnarray}
K^1{}_{10} &=&K^2{}_{20}=K^3{}_{30}=0,  \nonumber \\
K^1{}_{01} &=&K^2{}_{02}=K^3{}_{03}=h,  \nonumber \\
K^0{}_{11} &=&K^0{}_{22}=K^0{}_{22}={}a^2h,  \nonumber \\
K^1{}_{23} &=&K^2{}_{31}=K^3{}_{12}=-af,  \nonumber \\
K^1{}_{32} &=&K^2{}_{13}=K^3{}_{21}=af.
\end{eqnarray}

The non-vanishing components of the curvature $R^\rho {}_{\sigma \mu \nu }$
and the Ricci curvature $R{}_{\mu \nu }$ are

\begin{eqnarray}
R^0{}_{101} &=&R^0{}_{202}=R^0{}_{303}=a^2\left( \stackrel{{\bf \cdot }}{H}%
+H^2+Hh+\stackrel{\cdot }{h}\right) ,  \nonumber \\
R^0{}_{123} &=&-R^0{}_{213}=R^0{}_{312}=2a^3f\left( H+h\right) ,  \nonumber
\\
R^1{}_{203} &=&-R^1{}_{302}=R^2{}_{301}=-a\left( Hf+\stackrel{\cdot }{f}%
\right) ,  \nonumber \\
R^1{}_{212} &=&R^1{}_{313}=R^2{}_{323}=a^2\left( \left( H+h\right)
^2-f^2\right) ,
\end{eqnarray}
\begin{eqnarray}
R{}_{00} &=&-3\stackrel{\cdot }{H}-3\stackrel{\cdot }{h}-3H^2-3Hh,  \nonumber
\\
R{}_{11} &=&R{}_{22}=R{}_{33}=a^2\left( \stackrel{\cdot }{H}+\stackrel{\cdot
}{h}+3H^2+5Hh+2h^2-f^2\right) ,
\end{eqnarray}
\begin{equation}
R{}=6\stackrel{\cdot }{H}+6\stackrel{\cdot }{h}+12H^2+18Hh+6h^2-3f^2,
\end{equation}
where $H=\stackrel{\cdot }{a}\left( t\right) /a\left( t\right) $ is the
Hubble parameter. Using these results and supposing the matter source is a
fluid characterized by the energy density $\rho $, the pressure $p$ and the
spin $s_{IJ}{}^\mu $ we obtain four independent equations from (2) and (3):
\begin{eqnarray}
&&\left( H+h\right) ^2-f^2-\frac \rho 3  \nonumber \\
&&-\left( \beta +3\alpha \right) [4\left( \stackrel{\cdot }{H}+\stackrel{%
\cdot }{h}\right) ^2+8H\left( H+h\right) \left( \stackrel{\cdot }{H}+%
\stackrel{\cdot }{h}\right)  \nonumber \\
&&-4h\left( h+2H\right) \left( h+H\right) ^2+8\left( h+H\right) ^2f^2-4f^4]
\nonumber \\
&&+2\gamma \left( 3h^2+4f^2\right) =0,
\end{eqnarray}
\begin{eqnarray}
&&2\left( \stackrel{\cdot }{H}+\stackrel{\cdot }{h}\right)
+3H^2+4Hh+h^2-f^2-p  \nonumber \\
&&+\left( \beta +3\alpha \right) [4\left( \stackrel{\cdot }{H}+\stackrel{%
\cdot }{h}\right) ^2+8H\left( H+h\right) \left( \stackrel{\cdot }{H}+%
\stackrel{\cdot }{h}\right)  \nonumber \\
&&-\allowbreak 4h\left( h+2H\right) \left( h+H\right) ^2+8\left( h+H\right)
^2f^2+4f^4]  \nonumber \\
&&-2\gamma \left( 2\stackrel{\cdot }{h}+8Hh+h^2{}+4f^2\right) =0,
\end{eqnarray}
\begin{eqnarray}
&&\left( \beta +6\alpha \right) \left( \stackrel{\cdot \cdot }{H}+\stackrel{%
\cdot \cdot }{h}\right) +6\left( \beta +4\alpha \right) \left( H+h\right)
\stackrel{\cdot }{H}+\left( \allowbreak 5\beta +18\alpha \right) \left(
H+h\right) \stackrel{\cdot }{h}-4\left( \beta +3\alpha \right) f\stackrel{%
{\bf \cdot }}{f}  \nonumber \\
&&+3\left( \beta +4\alpha \right) hH^2+\left( \allowbreak 5\beta +18\alpha
\right) h^2H+2\left( \beta +3\alpha \right) h^3-2\left( \beta +3\alpha
\right) hf^2+\frac 14h+\frac 12s_{01}{}^1=0,
\end{eqnarray}
\begin{eqnarray}
&&f\{2\left( \beta +6\alpha \right) \left( \stackrel{\cdot }{H}+\stackrel{%
\cdot }{h}\right) +6\left( \beta +\allowbreak 4\alpha \right) H^2+2\left(
5\beta +18\alpha \right) Hh  \nonumber \\
&&+\left( \beta \ +3\alpha \right) \left( 4h^2-4f^2\right) -4\gamma +\frac 12%
\}-\frac 12s_{12}{}^3=0.
\end{eqnarray}
The system of the equations (14)--(17) has the similar structure as the
system of gravitational equations for homogeneous isotropic cosmological
models in [31] except the coefficients. However, it is the differences in
coefficients that make the system of the equations (14)--(17) easy to handle
and possible to obtain some exact or analytic solutions in several cases as
will be shown in the next sections.

The equations (14) and (15) can be written as
\begin{equation}
\stackrel{\cdot }{H}+\stackrel{\cdot }{h}=2\gamma \stackrel{\cdot }{h}%
-2H^2+\left( 8\gamma -3\right) Hh-\left( 2\gamma {}+1\right) h^2+f^2+\frac 16%
\left( \rho +3p\right) ,
\end{equation}
and
\begin{eqnarray}
&&\left( \beta +3\alpha \right) [-4\left( \stackrel{\cdot }{H}+\stackrel{%
\cdot }{h}\right) ^2-8\left( \stackrel{\cdot }{H}+\stackrel{\cdot }{h}%
\right) H\left( H+h\right)  \nonumber \\
&&+4h\left( h+2H\right) \left( h+H\right) ^2-8\left( h+H\right) ^2f^2+4f^4]
\nonumber \\
&&+\gamma \left( 6h^2+8f^2\right) +H^2+2Hh+h^2-f^2-\frac 13\rho =0.
\end{eqnarray}

Since the spin orientation of particles in ordinary matter is random, the
macroscopic spacetime average of the spin vanishes, we suppose $%
s_{IJ}{}^\lambda =0$, henceforth. Then, the equations (16), (17) become
\begin{eqnarray}
&&\left( \beta +6\alpha \right) \left( \stackrel{\cdot \cdot }{H}+\stackrel{%
\cdot \cdot }{h}\right) +6\left( \beta +4\alpha \right) \left( H+h\right)
\stackrel{\cdot }{H}+\left( \allowbreak 5\beta +18\alpha \right) \left(
H+h\right) \stackrel{\cdot }{h}-4\left( \beta +3\alpha \right) f\stackrel{%
\cdot }{f}  \nonumber \\
&&+3\left( \beta +4\alpha \right) hH^2+\left( \allowbreak 5\beta +18\alpha
\right) h^2H+2\left( \beta +3\alpha \right) h^3-2\left( \beta +3\alpha
\right) hf^2+\frac 14h=0,
\end{eqnarray}
and
\begin{eqnarray}
2 &&f\{2\left( \beta +6\alpha \right) \left( \stackrel{\cdot }{H}+\stackrel{%
\cdot }{h}\right) +6\left( \beta +\allowbreak 4\alpha \right) H^2  \nonumber
\\
&&+2\left( 5\beta +18\alpha \right) Hh+\left( \beta \ +3\alpha \right)
\left( 4h^2-4f^2\right) -4\gamma +\frac 12\}=0.
\end{eqnarray}
(21) has the solutions
\begin{equation}
f=0,
\end{equation}
and

\begin{eqnarray}
f^2 &=&\frac{\left( \beta +6\alpha \right) }{2\left( \beta \ +3\alpha
\right) }\left( \stackrel{\cdot }{H}+\stackrel{\cdot }{h}\right) +\frac{%
3\left( \beta +\allowbreak 4\alpha \right) }{2\left( \beta \ +3\alpha
\right) }H^2+\frac{\left( 5\beta +18\alpha \right) }{2\left( \beta \
+3\alpha \right) }Hh+h^2  \nonumber \\
&&-\frac \gamma {\left( \beta \ +3\alpha \right) }+\frac 1{8\left( \beta \
+3\alpha \right) }.
\end{eqnarray}

We will solve the equations (18-20) in the cases (22) and (23), respectively
in the next two sections.

\section{$\text{Exact de Sitter solutions with scalar torsion function}$}

In the case $f=0$, (18) and (19) can be written as
\begin{equation}
\stackrel{\cdot }{H}=\left( 2\gamma -1\right) \stackrel{\cdot }{h}%
-2H^2+\left( 8\gamma -3\right) Hh-\left( 2\gamma {}+1\right) h^2+\frac 16%
\left( \rho +3p\right) ,
\end{equation}
and
\begin{equation}
\left( \stackrel{\cdot }{h}+\left( 4H-h\right) h-\frac 1{2\gamma }\left(
H+h\right) ^2+\frac{\rho +3p}{12\gamma }\right) ^2-\frac{12\left( \beta
+3\alpha \right) \left( H+h\right) ^4+3\left( H+h\right) ^2+18h^2\gamma
-\rho }{48\gamma ^2\left( \beta +3\alpha \right) }=0,
\end{equation}
which have the solutions

\begin{equation}
\stackrel{\cdot }{h}=-\left( 4H-h\right) h+\frac 1{2\gamma }\left(
H+h\right) ^2-\frac{\rho +3p}{12\gamma }\pm \sqrt{\frac{12\left( \beta
+3\alpha \right) \left( H+h\right) ^4+3\left( H+h\right) ^2+18h^2\gamma
-\rho }{48\gamma ^2\left( \beta +3\alpha \right) }},
\end{equation}

\begin{eqnarray}
\stackrel{\cdot }{H} &=&-\frac 1{2\gamma }\left( H+h\right) ^2\allowbreak
-H^2+3Hh-h^2+\allowbreak \frac{\rho +3p}{12\gamma }  \nonumber \\
&&\pm \left( 2\gamma -1\right) \sqrt{\frac{12\left( \beta +3\alpha \right)
\left( H+h\right) ^4+3\left( H+h\right) ^2+18h^2\gamma -\rho }{48\gamma
^2\left( \beta +3\alpha \right) }},
\end{eqnarray}

Differentiating (24) gives
\begin{equation}
\stackrel{\cdot \cdot }{H}=\left( 2\gamma -1\right) \stackrel{\cdot \cdot }{h%
}-4H\stackrel{\cdot }{H}+\left( 8\gamma -3\right) \stackrel{\cdot }{H}%
h+\left( 8\gamma -3\right) H\stackrel{\cdot }{h}-2\left( 2\gamma {}+1\right)
h\stackrel{\cdot }{h}+\frac 16\left( \stackrel{\cdot }{\rho }+3\stackrel{%
\cdot }{p}\right) .
\end{equation}
The equation (20) has the form
\begin{eqnarray}
&&\left( \beta +6\alpha \right) \left( \stackrel{\cdot \cdot }{H}+\stackrel{%
\cdot \cdot }{h}\right) +6\left( \beta +4\alpha \right) \left( H+h\right)
\stackrel{\cdot }{H}+\left( \allowbreak 5\beta +18\alpha \right) \left(
H+h\right) \stackrel{\cdot }{h}  \nonumber \\
&&+3\left( \beta +4\alpha \right) hH^2+\left( \allowbreak 5\beta +18\alpha
\right) h^2H+2\left( \beta +3\alpha \right) h^3+\frac 14h=0.
\end{eqnarray}
(28) and (29) have the solutions
\begin{eqnarray}
\stackrel{\cdot \cdot }{H} &=&-\frac{\left( 48\alpha \gamma +12\beta \gamma
-2\beta \right) H+\left( 4\beta \gamma -3\beta -6\alpha \right) h}{2\gamma
\left( \beta +6\alpha \right) }\stackrel{\cdot }{H}  \nonumber \\
&&-\frac{\left( 2\beta \gamma -2\beta -12\alpha \gamma \right) H+\left(
14\beta \gamma +60\alpha \gamma -3\beta -6\alpha \right) h}{2\gamma \left(
\beta +6\alpha \right) }\stackrel{\cdot }{h}  \nonumber \\
&&-\frac{3\left( 2\gamma -1\right) \left( \beta +4\alpha \right) }{2\gamma
\left( \beta +6\alpha \right) }hH^2-\frac{\left( 2\gamma -1\right) \left(
5\beta +18\alpha \right) }{2\gamma \left( \beta +6\alpha \right) }h^2H-
\nonumber \\
&&\frac{\left( 2\gamma -1\right) \allowbreak \left( \beta +3\alpha \right) }{%
\gamma \left( \beta +6\alpha \right) }h^3-\frac{2\gamma -1}{8\gamma \left(
\beta +6\alpha \right) }h+\frac 1{12\gamma }\left( \stackrel{\cdot }{\rho }-3%
\stackrel{\cdot }{p}\right) ,
\end{eqnarray}
\begin{eqnarray}
\stackrel{\cdot \cdot }{h} &=&-\frac{2\beta H+\left( 8\gamma \beta +48\gamma
\alpha +3\beta +6\alpha \right) h}{2\gamma \left( \beta +6\alpha \right) }%
\stackrel{\cdot }{H}  \nonumber \\
&&-\frac{\left( 2\beta +8\gamma \beta +48\gamma \alpha \right) H+\left(
3\beta +6\alpha -4\gamma \beta -24\gamma \alpha \right) h}{2\gamma \left(
\beta +6\alpha \right) }\stackrel{\cdot }{h}  \nonumber \\
&&-\frac{3\left( \beta +4\alpha \right) h}{2\gamma \left( \beta +6\alpha
\right) }H^2-\frac{\allowbreak 5\beta +18\alpha }{2\gamma \left( \beta
+6\alpha \right) }h^2H-\frac{\beta +3\alpha }{\gamma \left( \beta +6\alpha
\right) }h^3  \nonumber \\
&&-\frac 1{8\gamma \left( \beta +6\alpha \right) }h-\frac 1{12\gamma }\left(
\stackrel{\cdot }{\rho }-3\stackrel{\cdot }{p}\right) .
\end{eqnarray}

Letting
\begin{equation}
\stackrel{\cdot }{H}=X,\stackrel{\cdot }{h}=Y,
\end{equation}
we have the dynamical system
\begin{eqnarray*}
\stackrel{\cdot }{X} &=&-\frac{\left( 48\alpha \gamma +12\beta \gamma
-2\beta \right) H+\left( 4\beta \gamma -3\beta -6\alpha \right) h}{2\gamma
\left( \beta +6\alpha \right) }X \\
&&-\frac{\left( 2\beta \gamma -2\beta -12\alpha \gamma \right) H+\left(
14\beta \gamma +60\alpha \gamma -3\beta -6\alpha \right) h}{2\gamma \left(
\beta +6\alpha \right) }Y\left( H,h\right) \\
&&-\frac{3\left( 2\gamma -1\right) \left( \beta +4\alpha \right) }{2\gamma
\left( \beta +6\alpha \right) }hH^2-\frac{\left( 2\gamma -1\right) \left(
5\beta +18\alpha \right) }{2\gamma \left( \beta +6\alpha \right) }h^2H- \\
&&\frac{\left( 2\gamma -1\right) \allowbreak \left( \beta +3\alpha \right) }{%
\gamma \left( \beta +6\alpha \right) }h^3-\frac{2\gamma -1}{8\gamma \left(
\beta +6\alpha \right) }h+\frac 1{12\gamma }\left( \stackrel{\cdot }{\rho }-3%
\stackrel{\cdot }{p}\right) ,
\end{eqnarray*}
\[
\stackrel{\cdot }{H}=X,
\]
\begin{eqnarray}
\stackrel{\cdot }{h} &=&Y\left( H,h\right) =\frac 1{2\gamma }\left(
H+h\right) ^2-\allowbreak \allowbreak \left( 4H-h\right) h-\frac 1{12\gamma }%
\left( \rho -3p\right)  \nonumber \\
&&\pm \sqrt{\allowbreak \frac{12\left( \beta +3\alpha \right) \left(
H+h\right) ^4+3\allowbreak \left( H+h\right) ^2+18\gamma h^2-\rho }{48\gamma
^2\left( \beta +3\alpha \right) }}.
\end{eqnarray}

The critical point equations consist of
\begin{eqnarray*}
&&-3\left( \beta +4\alpha \right) hH^2-\left( 5\beta +18\alpha \right) h^2H-
\\
&&2\allowbreak \left( \beta +3\alpha \right) h^3-\frac 14h+\frac{\beta
+6\alpha }{6\left( 2\gamma -1\right) }\left( \stackrel{\cdot }{\rho }-3%
\stackrel{\cdot }{p}\right) \\
&=&0,
\end{eqnarray*}
\[
X=0,
\]
\begin{eqnarray}
&&\left( H+h\right) ^2-2\gamma \allowbreak \allowbreak \left( 4H-h\right) h-%
\frac 16\left( \rho -3p\right)  \nonumber \\
&&\pm \sqrt{\allowbreak \frac{12\left( \beta +3\alpha \right) \left(
H+h\right) ^4+3\allowbreak \left( H+h\right) ^2+18\gamma h^2-\rho }{12\left(
\beta +3\alpha \right) }}=0.
\end{eqnarray}
In order to discuss the stability of the critical points we need to
calculate the matrix elements of the Jacobian:

\begin{eqnarray*}
\frac{\partial \stackrel{\cdot }{X}}{\partial X} &=&-\frac{\left( 48\alpha
\gamma +12\beta \gamma -2\beta \right) H+\left( 4\beta \gamma -3\beta
-6\alpha \right) h}{2\gamma \left( \beta +6\alpha \right) }, \\
\frac{\partial \stackrel{\cdot }{X}}{\partial H} &=&-\frac{\left( 48\alpha
\gamma +12\beta \gamma -2\beta \right) }{2\gamma \left( \beta +6\alpha
\right) }X-\frac{\left( 2\beta \gamma -2\beta -12\alpha \gamma \right) }{%
2\gamma \left( \beta +6\alpha \right) }Y\left( H,h\right) \\
&&-\frac{\left( 2\beta \gamma -2\beta -12\alpha \gamma \right) H+\left(
14\beta \gamma +60\alpha \gamma -3\beta -6\alpha \right) h}{2\gamma \left(
\beta +6\alpha \right) }\frac{\partial Y\left( H,h\right) }{\partial H} \\
&&-\frac{3\left( 2\gamma -1\right) \left( \beta +4\alpha \right) }{\gamma
\left( \beta +6\alpha \right) }hH-\frac{\left( 2\gamma -1\right) \left(
5\beta +18\alpha \right) }{2\gamma \left( \beta +6\alpha \right) }h^2,
\end{eqnarray*}

\begin{eqnarray*}
\frac{\partial \stackrel{\cdot }{X}}{\partial h} &=&-\frac{\left( 4\beta
\gamma -3\beta -6\alpha \right) }{2\gamma \left( \beta +6\alpha \right) }X-%
\frac{\left( 14\beta \gamma +60\alpha \gamma -3\beta -6\alpha \right) }{%
2\gamma \left( \beta +6\alpha \right) }Y\left( H,h\right) \\
&&-\frac{\left( 2\beta \gamma -2\beta -12\alpha \gamma \right) H+\left(
14\beta \gamma +60\alpha \gamma -3\beta -6\alpha \right) h}{2\gamma \left(
\beta +6\alpha \right) }\frac{\partial Y\left( H,h\right) }{\partial h} \\
&&-\frac{3\left( 2\gamma -1\right) \left( \beta +4\alpha \right) }{2\gamma
\left( \beta +6\alpha \right) }H^2-\frac{\left( 2\gamma -1\right) \left(
5\beta +18\alpha \right) }{\gamma \left( \beta +6\alpha \right) }hH \\
&&-\frac{3\left( 2\gamma -1\right) \allowbreak \left( \beta +3\alpha \right)
}{\gamma \left( \beta +6\alpha \right) }h^2-\frac{2\gamma -1}{8\gamma \left(
\beta +6\alpha \right) }
\end{eqnarray*}
\[
\frac{\partial \stackrel{\cdot }{H}}{\partial X}=1,\frac{\partial \stackrel{%
\cdot }{H}}{\partial H}=0,\frac{\partial \stackrel{\cdot }{H}}{\partial h}=0,%
\frac{\partial \stackrel{\cdot }{h}}{\partial X}=0,
\]

\begin{eqnarray*}
\frac{\partial \stackrel{\cdot }{h}}{\partial H} &=&\frac{\partial Y\left(
H,h\right) }{\partial H}=\frac 1\gamma H+\left( \frac 1\gamma -\allowbreak
\allowbreak 4\right) h \\
&&\pm \frac{8\left( \beta +3\alpha \right) \left( H+h\right) ^3+\allowbreak
H+h}{4\gamma \sqrt{\allowbreak 4\left( \beta +3\alpha \right) ^2\left(
H+h\right) ^4+\allowbreak \left( \beta +3\alpha \right) \left( H+h\right)
^2+6\left( \beta +3\alpha \right) \gamma h^2-\frac 13\left( \beta +3\alpha
\right) \rho }},
\end{eqnarray*}

\begin{eqnarray}
\frac{\partial \stackrel{\cdot }{h}}{\partial h} &=&\frac{\partial Y\left(
H,h\right) }{\partial h}=\left( \frac 1\gamma -\allowbreak \allowbreak
4\right) H+\left( \frac 1\gamma +2\right) h  \nonumber \\
&&\pm \frac{8\left( \beta +3\alpha \right) \left( H+h\right) ^3+H+h+6\gamma h%
}{4\gamma \sqrt{\allowbreak 4\left( \beta +3\alpha \right) ^2\left(
H+h\right) ^4+\left( \beta +3\alpha \right) \allowbreak \left( H+h\right)
^2+6\left( \beta +3\alpha \right) \gamma h^2-\frac 13\left( \beta +3\alpha
\right) \rho }}.
\end{eqnarray}
$\allowbreak $

In order to stress the role of the torsion as the source of the accelerating
expansion of the universe we concentrate on the vacuum solutions for some
special choices of the parameters $\alpha $, $\beta $, and $\gamma $. The
critical point equations (34) can be simplified and solved exactly when $%
\beta =4\alpha $ or $\beta =3\alpha $.

\subsection{$\label{0}$When $\protect\beta =-4\protect\alpha $}

In this case the gravitational Lagrangian is a special case of quadratic
curvature gravities [36] when the torsion vanishes.

In vacuum the dynamical system (33) becomes
\begin{eqnarray*}
\stackrel{\cdot }{X} &=&\left( 4h-\frac 2\gamma H-\frac 3{2\gamma }h\right)
X+\left( 5H-h-\frac 2\gamma H-\frac 3{2\gamma }h\right) Y\left( H,h\right) \\
&&+\frac{\left( 2\gamma -1\right) }{16\gamma \alpha }h\left( 8hH\alpha
+8\alpha h^2-1\right) ,
\end{eqnarray*}
\[
\stackrel{\cdot }{H}=X,
\]
\begin{eqnarray}
\stackrel{\cdot }{h} &=&Y\left( H,h\right) =\frac 1{2\gamma }\left(
H+h\right) ^2-\allowbreak \allowbreak \left( 4H-h\right) h  \nonumber \\
&&\pm \frac 1{4\gamma }\sqrt{\allowbreak \frac{4\alpha \left( H+h\right)
^4-\allowbreak \left( H+h\right) ^2-6\gamma h^2}\alpha },
\end{eqnarray}
and the critical point equations (34) become

\begin{equation}
h\left( 8hH\alpha +8\alpha h^2-1\right) =0,
\end{equation}
\begin{equation}
X=0,
\end{equation}
\begin{eqnarray}
&&\left( H+h\right) ^2-2\gamma \allowbreak \allowbreak \left( 4H-h\right) h
\nonumber \\
&&\pm \frac 12\sqrt{\allowbreak \frac{4\alpha \left( H+h\right)
^4-\allowbreak \left( H+h\right) ^2-6\gamma h^2}\alpha }=0.
\end{eqnarray}

The equation (37)
\[
h\left( 8hH\alpha +8\alpha h^2-1\right) =0,
\]
leads to
\[
h=0,
\]
or
\[
8hH\alpha +8\alpha h^2-1=0.
\]
Then we have two cases.

In the first case the solution
\begin{equation}
h=0,H=0,\stackrel{\cdot }{H}=X=0,
\end{equation}
corresponds to a static Minkowski spacetime.

In the second case, $h$ and $H$ satisfy the equations

\[
8hH\alpha +8\alpha h^2-1=0,
\]
and
\begin{eqnarray*}
&&\left( H+h\right) ^2-2\gamma \allowbreak \allowbreak \left( 4H-h\right) h
\\
&&\pm \frac 12\sqrt{\allowbreak \frac{4\alpha \left( H+h\right)
^4-\allowbreak \left( H+h\right) ^2-6\gamma h^2}\alpha }=0,
\end{eqnarray*}
which can be written as
\[
H=-h+\frac 1{8\alpha h},
\]
and

\begin{equation}
25600\alpha ^3h^6\gamma ^2+128\gamma \alpha ^2h^4\left( 3-40\gamma \right)
+16\gamma \alpha h^2\left( 16\gamma +5\right) -8\gamma +1=0.
\end{equation}

In order to obtain a concrete results we give some specific value of $\gamma
$.

When
\begin{equation}
\gamma =4,
\end{equation}
the equations (41) become
\[
H=-h+\frac 1{8\alpha h},
\]
$\allowbreak $and
\[
409600\alpha ^3h^6-80384\alpha ^2h^4+4416\alpha h^2-31=0,
\]
which have a real solution
\[
H=\frac{1.40162}{\sqrt{\alpha }},h=\frac{0.0841322}{\sqrt{\alpha }}.
\]
The Jacobian matrix of the dynamical system (36) given by (35) is

\[
M=\left(
\begin{array}{lll}
\allowbreak -\frac{0.\,39583}{\sqrt{\alpha }} & -\frac{2.\,093}\alpha &
\allowbreak -\frac{33.\,382}\alpha \\
1 & 0 & 0 \\
0 & -\frac{.\,33904}{\sqrt{\alpha }} & -\frac{5.\,4102}{\sqrt{\alpha }}%
\end{array}
\right) ,
\]
which has the eigenvalues: $-0.\,85371/\sqrt{\alpha },\allowbreak -4.\,951/%
\sqrt{\alpha },\allowbreak -1.\,3522\times 10^{-3}/\sqrt{\alpha }$. If $%
\alpha >0$, all the real parts of the eigenvalues are negative. This means
that the critical point
\begin{equation}
X_c=0,H_c=\frac{1.40162}{\sqrt{\alpha }},h_c=\frac{0.0841322}{\sqrt{\alpha }}%
,
\end{equation}
is asymptotically stable and then
\begin{equation}
\stackrel{\cdot }{H}=X=0,
\end{equation}
gives an asymptotically stable de Sitter solution.

By the same way we can compute for

\begin{equation}
\gamma =2,
\end{equation}
the dynamical system (36) has a real critical point

\begin{equation}
X_c=0,H_c=\frac{0.824837}{\sqrt{\alpha }},h_c=\frac{0.130802}{\sqrt{\alpha }}%
,
\end{equation}
the Jacobian has the eigenvalues: $-1.\,665/\sqrt{\alpha }+0.\,26071i/\sqrt{%
\alpha },\allowbreak -1.\,665/\sqrt{\alpha }-0.\,26071i/\sqrt{\alpha }%
,\allowbreak -3.\,6641\times 10^{-3}/\sqrt{\alpha }$.

For
\begin{equation}
\gamma =1,
\end{equation}
(36) has a real critical point
\begin{equation}
X_c=0,H_c=\frac{0.488003}{\sqrt{\alpha }},h_c=\frac{0.185576}{\sqrt{\alpha }}%
,
\end{equation}
the Jacobian has the eigenvalues: $-0.\,8932/\sqrt{\alpha }+0.\,97558i/\sqrt{%
\alpha },\allowbreak -0.\,8932/\sqrt{\alpha }-0.\,97558i/\sqrt{\alpha }%
,\allowbreak -0.0\,4017/\sqrt{\alpha }$.

For
\begin{equation}
\gamma =\frac 12,
\end{equation}
(36) has a real critical point
\begin{equation}
X_c=0,H_c=\frac{0.389146}{\sqrt{\alpha }},h_c=\frac{0.208985}{\sqrt{\alpha }}%
,
\end{equation}
the Jacobian has the eigenvalues: $-2.\,6618/\sqrt{\alpha },\allowbreak
-0.\,22666/\sqrt{\alpha },\allowbreak -3.\,5927\times 10^{-5}/\sqrt{\alpha }$%
. All the three critical points are asymptotically stable.

For
\begin{equation}
\gamma =\frac 14,
\end{equation}
(36) has a real critical point
\begin{equation}
X_c=0,H_c=\frac{0.488003}{\sqrt{\alpha }},h=\frac{0.185576}{\sqrt{\alpha }},
\end{equation}
the Jacobian has the eigenvalues: $-5.\,1634/\sqrt{\alpha },\allowbreak
-0.\,68424/\sqrt{\alpha },\allowbreak 3.\,8689\times 10^{-2}/\sqrt{\alpha }$%
. One of the eigenvalues is positive, which means that the critical point
(52) is unstable. These examples illustrate that the stability of the
critical points depends on $\gamma $, i.e. on the term $\gamma T{}^\mu
{}_{\nu \rho }T{}_\mu {}^{\nu \rho }$ in the action (1).

\subsection{When $\protect\beta =-3\protect\alpha $}

This corresponds to conformal (Weyl) gravity (resent see [37]) and Critical
Gravity [38] when the torsion vanishes.

In this case the system of the equations (14)--(17) has the form

\begin{equation}
3H^2+6Hh+3\left( 6\gamma +1\right) h^2+3\left( 8\gamma -1\right) f^2-\rho =0,
\end{equation}
\begin{equation}
2\stackrel{\cdot }{H}+2\left( 1-2\gamma \right) \stackrel{\cdot }{h}%
+3H^2+4\left( 1-4\gamma \right) Hh+\left( 1-2\gamma \right) h^2{}-\left(
1+8\gamma \right) f^2-p=0,
\end{equation}
\begin{equation}
3\alpha \left( \stackrel{\cdot \cdot }{H}+\stackrel{\cdot \cdot }{h}\right)
+6\alpha \left( H+h\right) \stackrel{\cdot }{H}+3\alpha \left( H+h\right)
\stackrel{\cdot }{h}+3\alpha Hh\left( H+h\right) +\frac 14h+\frac 12%
s_{01}{}^1=0,
\end{equation}
\begin{equation}
f\{6\alpha \left( \stackrel{\cdot }{H}+\stackrel{\cdot }{h}\right) +6\alpha
H^2+6\alpha Hh-4\gamma +\frac 12\}-\frac 12s_{12}{}^3=0.
\end{equation}

When $s_{IJ}{}^\mu =0$, the equation (56) leads to

\[
f=0,
\]
or

\begin{equation}
6\alpha \left( \stackrel{\cdot }{H}+\stackrel{\cdot }{h}\right) +6\alpha
H^2+6\alpha Hh-4\gamma +\frac 12=0.
\end{equation}
$\newline
\allowbreak $

We deal with only the first case $f=0$ in this section, then the equations
(53)--(55) can be written as$\allowbreak $%
\begin{equation}
h=\frac{-H\pm \sqrt{-6\gamma H^2+\frac 13\left( 1+6\gamma \right) \rho }}{%
1+6\gamma },
\end{equation}
\begin{equation}
\stackrel{\cdot }{h}=\frac 1{2\gamma -1}\stackrel{\cdot }{H}+\frac 3{2\left(
2\gamma -1\right) }H^2+\frac{2-8\gamma }{2\gamma -1}Hh-\frac 12h^2{}-\frac 1{%
2\left( 2\gamma -1\right) }p,
\end{equation}
and
\begin{equation}
\stackrel{\cdot \cdot }{H}+\stackrel{\cdot \cdot }{h}+2\left( H+h\right)
\stackrel{\cdot }{H}+\left( H+h\right) \stackrel{\cdot }{h}+hH\left(
H+h\right) +\frac 1{12\alpha }h=0,
\end{equation}
Differentiating (59) gives
\begin{equation}
\stackrel{\cdot \cdot }{h}=\frac 1{2\gamma -1}\stackrel{\cdot \cdot }{H}+%
\frac 3{2\gamma -1}H\stackrel{\cdot }{H}+\frac{2-8\gamma }{2\gamma -1}h%
\stackrel{\cdot }{H}+\frac{2-8\gamma }{2\gamma -1}H\stackrel{\cdot }{h}-h%
\stackrel{\cdot }{h}{}-\frac 1{2\left( 2\gamma -1\right) }\stackrel{\cdot }{p%
}.
\end{equation}
Substituting (58), (59) and (61) into (60) yields

\begin{eqnarray}
\stackrel{\cdot \cdot }{H} &=&-\allowbreak \frac{2\left( 12\gamma ^2-8\gamma
-3\right) }{\left( 2\gamma -1\right) \left( 1+6\gamma \right) }H\stackrel{%
\cdot }{H}+\frac 2{\left( 6\gamma +1\right) }P\stackrel{\cdot }{H}\allowbreak
\nonumber \\
&&+\frac{\left( 636\gamma ^2-48\gamma -7\right) }{2\left( 2\gamma -1\right)
\left( 1+6\gamma \right) ^2}H^3-\allowbreak \frac{\left( 6\gamma -1\right)
\left( 26\gamma -5\right) }{\left( 2\gamma -1\right) \left( 1+6\gamma
\right) ^2}H^2P-\allowbreak \allowbreak \frac{10\gamma -3}{4\gamma \left(
1+6\gamma \right) ^2}HP^2  \nonumber \\
&&+\frac{\left( 2\gamma -1\right) H}{24\gamma \alpha \left( 6\gamma
+1\right) }-\frac{\left( 2\gamma -1\right) P}{24\gamma \alpha \left( 6\gamma
+1\right) }-\allowbreak \frac{6\gamma -1}{4\gamma \left( 2\gamma -1\right) }%
Hp+\frac 1{4\gamma }\stackrel{\cdot }{p},
\end{eqnarray}
where
\begin{equation}
P=\pm \sqrt{-6\gamma H^2+\frac 13\left( 1+6\gamma \right) \rho }.
\end{equation}

Letting
\begin{equation}
\stackrel{\cdot }{H}=X,
\end{equation}
we have the dynamical system$\allowbreak $%
\begin{eqnarray}
\stackrel{\cdot }{H} &=&X,  \nonumber \\
\stackrel{\cdot }{X} &=&-\frac{2\left( 12\gamma ^2-8\gamma -3\right) }{%
\left( 2\gamma -1\right) \left( 6\gamma +1\right) }HX+\frac 2{\left( 6\gamma
+1\right) }PX  \nonumber \\
&&+\frac{\left( 636\gamma ^2-48\gamma -7\right) }{2\left( 2\gamma -1\right)
\left( 6\gamma +1\right) ^2}H^3-\allowbreak \frac{\left( 6\gamma -1\right)
\left( 26\gamma -5\right) }{\left( 2\gamma -1\right) \left( 6\gamma
+1\right) ^2}PH^2-\frac{10\gamma -3}{4\gamma \left( 6\gamma +1\right) ^2}P^2H
\nonumber \\
&&+\frac{\left( 2\gamma -1\right) H}{24\gamma \alpha \left( 6\gamma
+1\right) }-\frac{\left( 2\gamma -1\right) P}{24\gamma \alpha \left( 6\gamma
+1\right) }-\frac{6\gamma -1}{4\gamma \left( 2\gamma -1\right) }Hp+\frac 1{%
4\gamma }\stackrel{{\bf \cdot }}{p},
\end{eqnarray}
with the matrix elements of its Jacobian:
\begin{eqnarray}
\frac{\partial \stackrel{\cdot }{H}}{\partial H} &=&0,\frac{\partial
\stackrel{\cdot }{H}}{\partial X}=1,  \nonumber \\
\frac{\partial \stackrel{\cdot }{X}}{\partial X} &=&-\allowbreak \frac{%
2\left( 12\gamma ^2-8\gamma -3\right) }{\left( 2\gamma -1\right) \left(
1+6\gamma \right) }H+\frac 2{\left( 6\gamma +1\right) }P,  \nonumber \\
\frac{\partial \stackrel{\cdot }{X}}{\partial H} &=&-\allowbreak \frac{%
2\left( 12\gamma ^2-8\gamma -3\right) }{\left( 2\gamma -1\right) \left(
1+6\gamma \right) }X-\frac{12\gamma }{\left( 6\gamma +1\right) }\frac HP%
X\allowbreak  \nonumber \\
&&+\allowbreak \frac{6\gamma \left( 6\gamma -1\right) \left( 26\gamma
-5\right) }{\left( 2\gamma -1\right) \left( 1+6\gamma \right) ^2}\frac{H^3}P+%
\frac{3\left( 676\gamma ^2-80\gamma -1\right) }{2\left( 2\gamma -1\right)
\left( 1+6\gamma \right) ^2}\allowbreak H^2  \nonumber \\
&&-\allowbreak \frac{2\left( 6\gamma -1\right) \left( 26\gamma -5\right) }{%
\left( 2\gamma -1\right) \left( 1+6\gamma \right) ^2}HP-\allowbreak
\allowbreak \frac{10\gamma -3}{4\gamma \left( 1+6\gamma \right) ^2}P^2
\nonumber \\
&&+\frac{2\gamma -1}{24\gamma \alpha \left( 6\gamma +1\right) }+\frac{%
2\gamma -1}{4\alpha \left( 6\gamma +1\right) }\frac HP-\allowbreak \frac{%
6\gamma -1}{4\gamma \left( 2\gamma -1\right) }p.
\end{eqnarray}

The fixed point equations are

\begin{eqnarray}
X &=&0,  \nonumber \\
&&\frac{\left( 636\gamma ^2-48\gamma -7\right) }{2\left( 2\gamma -1\right)
\left( 6\gamma +1\right) ^2}H^3-\allowbreak \frac{\left( 6\gamma -1\right)
\left( 26\gamma -5\right) }{\left( 2\gamma -1\right) \left( 6\gamma
+1\right) ^2}PH^2-\frac{10\gamma -3}{4\gamma \left( 6\gamma +1\right) ^2}P^2H
\nonumber \\
&&+\frac{\left( 2\gamma -1\right) H}{24\gamma \alpha \left( 6\gamma
+1\right) }-\frac{\left( 2\gamma -1\right) P}{24\gamma \alpha \left( 6\gamma
+1\right) }-\frac{6\gamma -1}{4\gamma \left( 2\gamma -1\right) }Hp+\frac 1{%
4\gamma }\stackrel{\cdot }{p}  \nonumber \\
&=&0.
\end{eqnarray}

Using (63) we obtain the equation
\begin{eqnarray}
&&\frac{348\gamma ^2-48\gamma +1}{\left( 2\gamma -1\right) \left( 1+6\gamma
\right) ^2}H^3+\frac{\left( 2\gamma -1\right) H}{24\gamma \alpha \left(
6\gamma +1\right) }-\left( \allowbreak \frac{\left( 6\gamma -1\right) \left(
26\gamma -5\right) }{\left( 2\gamma -1\right) \left( 6\gamma +1\right) ^2}%
H^2+\frac{\left( 2\gamma -1\right) }{24\gamma \alpha \left( 6\gamma
+1\right) }\right) P  \nonumber \\
&&-\frac{10\gamma -3}{12\gamma \left( 6\gamma +1\right) }H\rho -\frac{%
6\gamma -1}{4\gamma \left( 2\gamma -1\right) }Hp+\frac 1{4\gamma }\stackrel{%
\cdot }{p}  \nonumber \\
&=&0.
\end{eqnarray}

In vacuum
\begin{equation}
\rho =p=0,P=\nu H,\nu =\pm \sqrt{-6\gamma },
\end{equation}
The fixed point equation (68) becomes
\begin{equation}
\frac{348\gamma ^2-48\gamma +1}{\left( 2\gamma -1\right) \left( 1+6\gamma
\right) ^2}H^3+\frac{\left( 2\gamma -1\right) H}{24\gamma \alpha \left(
6\gamma +1\right) }-\left( \allowbreak \frac{\left( 6\gamma -1\right) \left(
26\gamma -5\right) }{\left( 2\gamma -1\right) \left( 6\gamma +1\right) ^2}%
H^2+\frac{2\gamma -1}{24\gamma \alpha \left( 6\gamma +1\right) }\right) \nu
H=0,
\end{equation}
which leads to
\begin{equation}
H=0,
\end{equation}
or
\begin{equation}
H^2=-\frac{\left( \nu -1\right) \left( 2\gamma -1\right) ^2\left( 1+6\gamma
\right) }{24\gamma \alpha \left( \left( 156\gamma ^2-56\gamma +5\right) \nu
-348\gamma ^2+48\gamma -1\right) }.
\end{equation}

In the first cas, we have
\begin{equation}
\stackrel{\cdot }{H}=X=0,H=0,h=0,
\end{equation}
which correspond to a static Minkowski solution.

In the second case
\begin{equation}
H^2=-\frac{\left( \nu -1\right) \left( 2\gamma -1\right) ^2\left( 1+6\gamma
\right) }{24\gamma \alpha \left( \left( 156\gamma ^2-56\gamma +5\right) \nu
-348\gamma ^2+48\gamma -1\right) },
\end{equation}
according to (66) the Jacobian matrix has the form
\begin{equation}
M=\left(
\begin{array}{ll}
0 & 1 \\
b & c%
\end{array}
\right) ,
\end{equation}
with
\begin{eqnarray}
b &=&\allowbreak \frac{6\gamma \left( 6\gamma -1\right) \left( 26\gamma
-5\right) }{\left( 2\gamma -1\right) \left( 1+6\gamma \right) ^2}\frac{H^2}%
\nu +\frac{3\left( 676\gamma ^2-80\gamma -1\right) }{2\left( 2\gamma
-1\right) \left( 1+6\gamma \right) ^2}\allowbreak H^2  \nonumber \\
&&-\allowbreak \frac{2\left( 6\gamma -1\right) \left( 26\gamma -5\right) }{%
\left( 2\gamma -1\right) \left( 1+6\gamma \right) ^2}\nu H^2-\allowbreak
\allowbreak \frac{10\gamma -3}{4\gamma \left( 1+6\gamma \right) ^2}\nu ^2H^2
\nonumber \\
&&+\frac{2\gamma -1}{24\gamma \alpha \left( 6\gamma +1\right) }+\frac{%
2\gamma -1}{4\alpha \left( 6\gamma +1\right) \nu },  \nonumber \\
c &=&-\frac{2\left( 12\gamma ^2-8\gamma -3-2\nu \gamma +\nu \right) }{\left(
2\gamma -1\right) \left( 1+6\gamma \right) }H.
\end{eqnarray}
and has the eigenvalues: $\frac 12c+\frac 12\sqrt{\left( c^2+4b\right) }%
,\allowbreak \frac 12c-\frac 12\sqrt{\left( c^2+4b\right) }$.

If
\[
\gamma =-\frac 1{24},
\]
(69), (74) and (76) give, separately,
\[
\nu =-\frac 12,
\]
\[
H^2=\frac{169}{948\alpha },
\]
$\allowbreak $
\[
b=-\frac{408901}{30336\alpha },c=-\frac{13}7\sqrt{\frac{169}{948\alpha }}.
\]
Then the Jacobian matrix has the eigenvalues:

\begin{eqnarray*}
\frac 12c+\frac 12\sqrt{\left( c^2+4b\right) } &=&-\frac{13}{14\sqrt{\alpha }%
}\sqrt{\frac{169}{948}}+\frac i{2\sqrt{\alpha }}\sqrt{-\frac{19807661}{371616%
}},\allowbreak \\
\frac 12c-\frac 12\sqrt{\left( c^2+4b\right) } &=&-\frac{13}{14\sqrt{\alpha }%
}\sqrt{\frac{169}{948}}-\frac i{2\sqrt{\alpha }}\sqrt{-\frac{19807661}{371616%
}}.
\end{eqnarray*}
If $\alpha >0$, all the real parts of the two eigenvalues are negative. This
means that the critical point
\begin{equation}
X_c=0,H_c=\sqrt{\frac{169}{948\alpha }},
\end{equation}
is asymptotically stable and then
\begin{equation}
\stackrel{\cdot }{H}=X=0,
\end{equation}
gives an asymptotically stable de Sitter solution. (58) gives
\[
h=-\frac{13}{711}\frac{\sqrt{237}}{\sqrt{\alpha }},
\]
or

\[
h=-\frac{13}{237}\frac{\sqrt{237}}{\sqrt{\alpha }}.
\]

We have seen that by appropriate choices of $\gamma $, we can obtain
asymptotically stable de Sitter solutions in both cases, $\beta =4\alpha $
and $\beta =3\alpha $, though the cosmological equations have different
structure. This means that the structure of the cosmological equations
depends on $\beta $, while the stability of the solutions depends on $\gamma
.$

In all of the solutions obtained the torsion function $h$ does not vanish in
vacuum, which means that the torsion is an intrinsic geometric nature of the
spacetime. It is the torsion that causes the accelerating expansion of the
universe in vacuum.

\section{$\text{Analytic solutions with pseudoscalar torsion function}$}

Differentiating (23) gives

\begin{equation}
f\stackrel{\cdot }{f}=\frac{\beta +6\alpha }{4\left( \beta +3\alpha \right) }%
\left( \stackrel{\cdot \cdot }{H}+\stackrel{\cdot \cdot }{h}\right) +\frac{%
3\left( \beta +\allowbreak 4\alpha \right) }{2\left( \beta +3\alpha \right) }%
H\stackrel{\cdot }{H}+\frac{5\beta +18\alpha }{4\left( \beta +3\alpha
\right) }\stackrel{\cdot }{H}h+\frac{5\beta +18\alpha }{4\left( \beta
+3\alpha \right) }H\stackrel{\cdot }{h}+h\stackrel{\cdot }{h}.
\end{equation}
Substituting (23) and (79) into (20) gives

\begin{equation}
h=0.
\end{equation}
Then the equations (18), (19) and (23) become
\begin{equation}
\stackrel{\cdot }{H}=-2H^2+f^2+\frac 16\left( \rho +3p\right) ,
\end{equation}
\begin{equation}
\left( \beta +3\alpha \right) [-4\stackrel{\cdot }{H}^2-8\stackrel{\cdot }{H}%
H^2-8H^2f^2+4f^4]+\left( 8\gamma -1\right) f^2+H^2-\frac 13\rho =0,
\end{equation}
\begin{equation}
f^2=\frac{\left( \beta +6\alpha \right) }{2\left( \beta \ +3\alpha \right) }%
\stackrel{\cdot }{H}+\frac{3\left( \beta +\allowbreak 4\alpha \right) }{%
2\left( \beta \ +3\alpha \right) }H^2-\frac \gamma {\left( \beta \ +3\alpha
\right) }+\frac 1{8\left( \beta \ +3\alpha \right) }.
\end{equation}
They have the solutions
\begin{equation}
H^2\allowbreak =\frac{\left( 8\gamma -1\right) ^2}{32\gamma \beta }+\frac 1{%
24\gamma }\rho -\frac{\left( 8\gamma -1\right) \left( \beta +4\alpha \right)
}{16\gamma \beta }\left( \rho +3p\right) +\frac{\left( \beta +3\alpha
\right) \left( \beta +4\alpha \right) }{24\gamma \beta }\left( \rho
+3p\right) ^2,
\end{equation}

\begin{equation}
\stackrel{{\bf \cdot }}{H}=\allowbreak -\frac{\left( 8\gamma -1\right)
\left( 16\gamma -1\right) }{32\gamma \beta }-\frac 1{24\gamma }\rho +\frac{%
40\gamma \beta +144\alpha \gamma -3\beta -12\alpha }{48\gamma \beta }\left(
\rho +3p\right) -\frac{\left( \beta +3\alpha \right) \left( \beta +4\alpha
\right) }{24\gamma \beta }\left( \rho +3p\right) ^2
\end{equation}

\begin{equation}
f^2=\allowbreak \frac{1-8\gamma }{32\gamma \beta }+\frac 1{24\gamma }\rho -%
\frac{16\gamma \left( \beta +3\alpha \right) -3\left( \beta +4\alpha \right)
}{48\gamma \beta }\left( \rho +3p\right) +\frac{\left( \beta +3\alpha
\right) \left( \beta +4\alpha \right) }{24\gamma \beta }\left( \rho
+3p\right) ^2.
\end{equation}
$\allowbreak $

The equations (84) and (85) ply the roles of the Friedmann equation and the
Raychaudhuri equation in General Relativity. The equation (86) indicates
that even in vacuum the spacetime possesses the torsion $f=\sqrt{\allowbreak
\frac{1-8\gamma }{32\gamma \beta }}$, which has been found in [32]. Hence
the conception of the vacuum as physical notion is changed essentially.
Instead of it as passive receptacle of physical objects and processes, the
vacuum assumes a dynamical properties as a gravitating object. The
combination of (84) and (85) yields the acceleration equation

\begin{equation}
\frac{\stackrel{\cdot \cdot }{a}}a=-\frac{8\gamma -1}{4\beta }+\frac{\beta \
+3\alpha }{3\beta }\left( \rho +3p\right) .
\end{equation}

Letting
\[
\beta =n\alpha ,
\]
we have
\begin{equation}
\frac{\stackrel{\cdot \cdot }{a}}a=\frac{1-8\gamma }{4n\alpha }+\frac{n\ +3}{%
3n}\left( \rho +3p\right) .
\end{equation}
Some important consequences can be obtained from (88):

i) The term $\frac{1-8\gamma }{4n\alpha }$ plies the role of the
cosmological constant, which agrees with the result in [31]. If $\frac{%
1-8\gamma }{4n\alpha }>0,$ $\rho =p=0,$ then $\stackrel{\cdot \cdot }{a}>0$,
the {\em acceleration} of cosmological expansion {\em acquires the vacuum
origin}.

ii) If
\begin{equation}
n>0,\text{ or }n<-3
\end{equation}
$\rho +3p$ accelerates the expansion of the universe. If
\begin{equation}
-3<n<0,
\end{equation}
$\rho +3p$ decelerates the expansion of the universe. Especially, when $%
n=-2,\gamma =1/8$, (88) becomes the acceleration equation in general
relativity. In other words, the latter is only a special case of the former.

iii) If
\begin{equation}
n>0,\gamma >\frac 18,\text{ or }n<-3,\gamma <\frac 18,
\end{equation}
the universe can undergo a phase transformation from an accelerating to a
decelerating expansion.

iv) If
\begin{equation}
-3<n<0,\gamma >\frac 18,
\end{equation}
the universe can undergo a phase transformation from a decelerating to an
accelerating expansion.

We find this picture very appealing and physical since it seems to indicate
that in metric-affine gravity as matter tells spacetime how to curve, matter
will also tell spacetime how to twirl.

\section{$\text{Analytic solutions in the case $\protect\gamma =0$}$}

In the last two sections we have seen that the coefficient $\gamma $ plies a
central role in determining the behavior of the scale factor and the
evolution of the universe. In order to investigate this point thoroughly, we
discuss a extreme case, $\gamma =0$. In this case the equations (14)--(17)
take the form

\begin{eqnarray}
&&\left( H+h\right) ^2-f^2-\frac \rho 3  \nonumber \\
&&+\left( \beta +3\alpha \right) [-4\left( \stackrel{\cdot }{H}+\stackrel{%
\cdot }{h}\right) ^2-8H\left( H+h\right) \left( \stackrel{\cdot }{H}+%
\stackrel{\cdot }{h}\right)  \nonumber \\
&&+4h\left( h+2H\right) \left( h+H\right) ^2-8\left( h+H\right)
^2f^2+4f^4]=0,
\end{eqnarray}
\begin{eqnarray}
&&2\left( \stackrel{\cdot }{H}+\stackrel{\cdot }{h}\right)
+3H^2+4Hh+h^2-f^2-p  \nonumber \\
&&-\left( \beta +3\alpha \right) [-4\left( \stackrel{\cdot }{H}+\stackrel{%
\cdot }{h}\right) ^2-8\left( H^2+Hh\right) \left( \stackrel{\cdot }{H}+%
\stackrel{\cdot }{h}\right)  \nonumber \\
&&+\allowbreak 4h\left( h+2H\right) \left( h+H\right) ^2-8\left( h+H\right)
^2f^2+4f^4]=0,
\end{eqnarray}
\begin{eqnarray}
&&\left( \beta +6\alpha \right) \left( \stackrel{\cdot \cdot }{H}+\stackrel{%
\cdot \cdot }{h}\right) +6\left( \beta +4\alpha \right) \left( H+h\right)
\stackrel{\cdot }{H}+\left( \allowbreak 5\beta +18\alpha \right) \left(
H+h\right) \stackrel{\cdot }{h}-4\left( \beta +3\alpha \right) f\stackrel{%
\cdot }{f}  \nonumber \\
&&+3\left( \beta +4\alpha \right) hH^2+\left( \allowbreak 5\beta +18\alpha
\right) h^2H+2\left( \beta +3\alpha \right) h^3-2\left( \beta +3\alpha
\right) hf^2+\frac 14h+\frac 12s_{01}{}^1=0,
\end{eqnarray}
\begin{eqnarray}
&&f\{2\left( \beta +6\alpha \right) \left( \stackrel{\cdot }{H}+\stackrel{%
\cdot }{h}\right) +6\left( \beta +\allowbreak 4\alpha \right) H^2  \nonumber
\\
&&+2\left( 5\beta +18\alpha \right) Hh+\left( \beta \ +3\alpha \right)
\left( 4h^2-4f^2\right) +\frac 12\}-\frac 12s_{12}{}^3=0.
\end{eqnarray}
(93) and (94) can be written as
\begin{equation}
\stackrel{\cdot }{H}+\stackrel{\cdot }{h}=-2H^2-3Hh-h^2+f^2+\frac 16\left(
\rho +3p\right) ,
\end{equation}
\begin{eqnarray}
&&3\left( h+H\right) ^2-3f^2-\rho  \nonumber \\
&&+4\left( \beta +3\alpha \right) \left( \left( H+h\right) ^2-f^2\right)
\left( \rho +3p\right) -\frac 13\left( \beta +3\alpha \right) \left( \rho
+3p\right) ^2  \nonumber \\
&=&0,
\end{eqnarray}
Differentiating (97) gives
\begin{equation}
\stackrel{\cdot \cdot }{H}+\stackrel{\cdot \cdot }{h}=-4H\stackrel{\cdot }{H}%
-3h\stackrel{\cdot }{H}-3H\stackrel{\cdot }{h}-2h\stackrel{\cdot }{h}+2f%
\stackrel{\cdot }{f}+\frac 16\left( \stackrel{\cdot }{\rho }+3\stackrel{%
\cdot }{p}\right) .
\end{equation}
Substituting (97) and (99) into (95) and (96) yields

\begin{eqnarray}
&&-\beta \left( h+4H\right) \left( h+H\right) ^2+\allowbreak \beta \left(
h+2H\right) f^2-\allowbreak 2\beta f\stackrel{\cdot }{f}+\frac 14%
h\allowbreak +\frac 16\left( 2\beta H+6\alpha h+3\beta h\right) \left( \rho
+3p\right)  \nonumber \\
&&+\allowbreak \frac 16\left( \beta +6\alpha \right) \left( \stackrel{\cdot }%
{\rho }+3\stackrel{\cdot }{p}\right) +\frac 12s_{01}  \nonumber \\
&=&0.
\end{eqnarray}
and
\begin{equation}
f\{2\beta \left( h+H\right) ^2-2\beta f^2+\frac 13\left( \beta +6\alpha
\right) \left( \rho +3p\right) +\frac 12\}-\frac 12s_{12}{}^3=0.
\end{equation}
If
\[
s_{IJ}{}^\mu =0,
\]
(101) leads to
\[
f=0,
\]
or
\begin{equation}
2\beta \left( h+H\right) ^2-2\beta f^2+\frac 13\left( \beta +6\alpha \right)
\left( \rho +3p\right) +\frac 12=0.
\end{equation}

\subsection{When $f=0$}

The equations (98) and (100) become
\begin{equation}
\left( H+h\right) ^2=\frac{\frac 13\left( \beta +3\alpha \right) \left( \rho
+3p\right) ^2+\rho }{4\left( \beta +3\alpha \right) \left( 3p+\rho \right) +3%
},
\end{equation}
and
\begin{eqnarray}
&&\frac 14h+\frac 16\allowbreak \left( 2\beta H+3\beta h+6\alpha h\right)
\left( \rho +3p\right) \allowbreak +\left( \beta +6\alpha \right) \left(
\stackrel{\cdot }{\rho }+3\stackrel{\cdot }{p}\right)  \nonumber \\
&&-\beta \left( h+4H\right) \left( H+h\right) ^2=0,
\end{eqnarray}
which have the solutions
\begin{eqnarray}
H &=&\frac{\frac 34+\allowbreak \frac 32\beta \left( \rho +5p\right)
+\allowbreak 6\alpha \left( \rho +3p\right) +\allowbreak \frac 13\left(
\beta +3\alpha \right) \left( 5\beta +12\alpha \right) \left( \rho
+3p\right) ^2}{\frac 34+\allowbreak \frac 92\beta \left( \rho +p\right)
+\allowbreak 6\alpha \left( \rho +3p\right) +\allowbreak \frac 13\left(
\beta +3\alpha \right) \left( 5\beta +12\alpha \right) \left( \rho
+3p\right) ^2}J \\
&&+\frac{4\left( \beta +3\alpha \right) \left( \beta +6\alpha \right) \left(
\rho +3p\right) +\allowbreak 3\left( \beta +6\alpha \right) }{\frac 34%
+\allowbreak \frac 92\beta \left( \rho +p\right) +\allowbreak 6\alpha \left(
\rho +3p\right) +\allowbreak \frac 13\left( \beta +3\alpha \right) \left(
5\beta +12\alpha \right) \left( \rho +3p\right) ^2}\left( \stackrel{\cdot }{%
\rho }+3\stackrel{\cdot }{p}\right) ,  \nonumber
\end{eqnarray}

\begin{eqnarray}
h &=&\frac{3\beta \left( \rho -p\right) }{\frac 34+\allowbreak \frac 92\beta
\left( \rho +p\right) +\allowbreak 6\alpha \left( \rho +3p\right)
+\allowbreak \frac 13\left( \beta +3\alpha \right) \left( 5\beta +12\alpha
\right) \left( \rho +3p\right) ^2}J  \nonumber \\
&&-\frac{4\left( \beta +3\alpha \right) \left( \beta +6\alpha \right) \left(
\rho +3p\right) +\allowbreak 3\left( \beta +6\alpha \right) }{\frac 34%
+\allowbreak \frac 92\beta \left( \rho +p\right) +\allowbreak 6\alpha \left(
\rho +3p\right) +\allowbreak \frac 13\left( \beta +3\alpha \right) \left(
5\beta +12\alpha \right) \left( \rho +3p\right) ^2}\left( \stackrel{\cdot }{%
\rho }+3\stackrel{\cdot }{p}\right) ,
\end{eqnarray}
with
\begin{equation}
J=\pm \sqrt{\frac{\frac 13\left( \beta +3\alpha \right) \left( \rho
+3p\right) ^2+\rho }{4\left( \beta +3\alpha \right) \left( 3p+\rho \right) +3%
}}.
\end{equation}
The equation (97) now becomes
\begin{equation}
\stackrel{\cdot }{H}+\stackrel{\cdot }{h}=-\left( 2H+h\right) \left(
H+h\right) +\frac 16\left( \rho +3p\right) .
\end{equation}

Letting
\begin{equation}
\beta =n\alpha ,p=w\rho ,
\end{equation}
(105), (106) and (107) can be written as

\begin{equation}
H=\frac{\left( \frac 34+\allowbreak \frac 32A\alpha \rho +\frac 13B\alpha
^2\rho ^2\right) J+\left( 4D\alpha \rho +\allowbreak G\right) \alpha \left(
1+3w\right) \stackrel{\cdot }{\rho }}{\frac 34+\allowbreak \frac 32C\alpha
\rho +\frac 13B\alpha ^2\rho ^2},
\end{equation}

\begin{equation}
h=\frac{3n\left( 1-w\right) \alpha \rho J-\left( 4D\alpha \rho +G\right)
\alpha \left( 1+3w\right) \stackrel{\cdot }{\rho }}{\frac 34+\frac 32C\alpha
\rho +\allowbreak \frac 13B\alpha ^2\rho ^2},
\end{equation}
\begin{equation}
J=\pm \sqrt{\frac{\frac 13E\alpha \rho +1}{4F\alpha \rho +3}\rho },
\end{equation}
where
\begin{eqnarray}
A &=&n+4+\left( 5n+12\right) w,B=\left( n+3\right) \left( 5n+12\right)
\left( 1+3w\right) ^2,  \nonumber \\
C &=&3n+4+\left( 3n+12\right) w,D=\left( n+3\right) \left( n+6\right) \left(
1+3w\right) ,  \nonumber \\
E &=&\left( n+3\right) \left( 1+3w\right) ^2,F=\left( n+3\right) \left(
3w+1\right) ,G=3\left( n+6\right) .
\end{eqnarray}
The equation (108) gives the acceleration equation

\begin{equation}
\frac{\stackrel{\cdot \cdot }{a}}a=-\stackrel{\cdot }{h}\allowbreak -\left(
H+h\right) ^2-hH+\frac 16\left( 1+3w\right) \rho ,
\end{equation}

Let us consider two special cases.

i) For the early universe,
\[
\alpha \rho \gg 1,
\]
we compute using (110)-- (114) and obtain approximately
\begin{equation}
\frac{\stackrel{\cdot \cdot }{a}}a=\allowbreak \frac 1{12}\left( 1+3w\right)
\rho .
\end{equation}
This represents an inflation universe.

ii) For the later epoch
\[
\alpha \rho \ll 1,
\]
we have
\begin{equation}
\frac{\stackrel{\cdot \cdot }{a}}a=-\allowbreak \frac 16\rho \left(
1-3w\right) .
\end{equation}
This represents a uniformly expanding universe if $w=1/3$ (radiation epoch),
or an decelerating universe if $w<1/3$ (matter epoch).

\subsection{when $f\neq 0$}

The function $f$ satisfies the equation (102), which yields
\begin{equation}
f^2=\left( H+h\right) ^2+\frac{\beta +6\alpha }{6\beta }\left( \rho
+3p\right) +\frac 1{4\beta },
\end{equation}
and
\[
f\stackrel{\cdot }{f}=\left( H+h\right) \left( \stackrel{\cdot }{H}+\stackrel%
{\cdot }{h}\right) +\frac{\beta +6\alpha }{12\beta }\left( \stackrel{\cdot }{%
\rho }+3\stackrel{\cdot }{p}\right) .
\]
Using (97) we have
\begin{equation}
\stackrel{\cdot }{H}+\stackrel{\cdot }{h}=-H\left( H+h\right) +\frac{\beta
+3\alpha }{3\beta }\left( \rho +3p\right) +\frac 1{4\beta },
\end{equation}
\begin{equation}
f\stackrel{\cdot }{f}=-H\left( H+h\right) ^2+\frac{\beta +3\alpha }{3\beta }%
\left( H+h\right) \left( \rho +3p\right) +\frac 1{4\beta }\left( H+h\right) +%
\frac{\beta +6\alpha }{12\beta }\left( \stackrel{\cdot }{\rho }+3\stackrel{%
\cdot }{p}\right) ,
\end{equation}
substituting in to (98) and (100) we have
\begin{equation}
-\frac{\beta +4\alpha }{3\beta }\left( \rho +3p\right) ^2-\frac{\beta
+4\alpha }{2\beta \left( \beta +3\alpha \right) }\left( \rho +3p\right) -%
\frac \rho {3\left( \beta +3\alpha \right) }-\frac 1{4\beta \left( \beta
+3\alpha \right) }=0,
\end{equation}
\begin{equation}
0=0.
\end{equation}
So the field equations have no definite solution.

The results obtain above indicates that if $\gamma =0$ in both cases, $f=0$,
and $f\neq 0$, there exists no solution describing an accelerating universe.
In other words, the term $\gamma T{}^\mu {}_{\nu \rho }T{}_\mu {}^{\nu \rho
} $ is necessary to the existence of the solutions describing an
accelerating universe.

\section{$\text{Conclusions}$}

Quadratic theories of gravity described by the Lagrangian $R+\alpha
R^2+\beta R_{\mu \nu }R^{\mu \nu }$ have been studied in many works in
supergravity, quantum gravity, string theory and M-theory. However, the
cosmology in these theories has not been explored extensively, especially,
when the torsion of the spacetime is considered. In this paper we show that
by only allowing the connection to be asymmetrical and adding a term $\gamma
T{}^\mu {}_{\nu \rho }T{}_\mu {}^{\nu \rho }$ to the Lagrangian $R+\alpha
R^2+\beta R_{\mu \nu }R^{\mu \nu }$ some meaningful cosmological solutions
can be obtained. These solutions provide several possible explanations to
the acceleration of the cosmological expansion without a cosmological
constant or dark energy. One can find that although the field equation (2)
returns to Einstein's equation when $\alpha =\beta =\gamma =0$, the
cosmological equations (18-21) are essentially different from the Friedmann
equation and the Raychaudhuri equation and then give different description
to the evolution of the universe. The acceleration equation of the universe
in general relativity is only a special case of the equation (87). These
equations involving higher-derivatives can be solved by appropriate choice
of $\beta $ and $\gamma $. Not only numbers of asymptotically stable de
Sitter solutions expressed by critical points of a dynamical system but also
exact analytic solutions are obtained. These solutions indicate that the
terms $\beta R_{\mu \nu }R^{\mu \nu }$ and $\gamma T{}^\mu {}_{\nu \rho
}T{}_\mu {}^{\nu \rho }$ ply the different roles: the former determines the
structure of the equations while the latter determines the behavior and the
stability of the solutions. To construct a model of cosmic acceleration the
Lagrangian $R+\alpha R^2+\beta R_{\mu \nu }R^{\mu \nu }+$ $\gamma T{}^\mu
{}_{\nu \rho }T{}_\mu {}^{\nu \rho }$ is sufficient and necessary.

Owing to the solutions obtained some conceptions have to be\ changed
essentially. According to these solutions, even in vacuum the spacetime can
possesses torsion and curvature. Therefore, instead of vacuum as passive
receptacle of physical objects and processes, the vacuum assumes a dynamical
property as a gravitating object. It is the torsion of the spacetime that
causes the acceleration of the cosmological expansion in vacuum. Both the
torsion and the accelerating expansion possess geometrical nature and do not
invoke any matter origin. Furthermore, the energy and pressure of the
ordinary matter can produce the torsion of the spacetime and cause either
the deceleration or the acceleration of the cosmological expansion depending
on choices of $\beta $ and $\gamma $.


\begin{thebibliography}{*}

 \bibitem{} Capozziello S and De Laurentis M, arXiv:1108.6266 [gr-qc]

\bibitem{} De Felice A and Tsujikawa S 2010 {\it Living Rev. Rel}. {\bf 13} 3

\bibitem{} Sotiriou T P and Faraoni V 2010 {\it Rev. Mod. Phys}. {\bf 82} 451

\bibitem{} Nojiri S and Odintsov S D 2011 {\it Phys. Rept.} {\bf 505} 59

\bibitem{} Sotiriou T P 2009 {\it J. Phys. Conf. Ser.}{\bf 189} 012039
(arXiv:0810.5594 [gr-qc])

\bibitem{} Nojiri S and Odintsov S D 2007 {\it Int. J. Geom. Meth. Mod. Phys}%
. {\bf 4} 115

\bibitem{} Buchdahl H A 1970 {\it Mon. Not. Roy. Ast. Soc}. {\bf 150} 1

\bibitem{} Capozziello S and Francaviglia M 2008 {\it Gen. Rel. Grav.} {\bf %
40} 357

\bibitem{} Sotiriou T P 2009 {\it Class. Quant. Grav}. {\bf 26} 152001

Vitagliano V, Sotiriou T P and Liberati S 2010 {\it Phys. Rev}. D{\bf 82}
084007 {\bf \ }

\bibitem{} Barragan C and Olmo G J 2010{\bf \ }{\it Phys. Rev}. D{\bf 82}
084015

Olmo G J, Sanchis-Alepuz H and Tripathi S 2009 {\it Phys. Rev}. D{\bf 80}
024013 {\bf \ }

Barragan C, Olmo G J and Sanchis-Alepuz H 2009 {\it Phys. Rev}. D{\bf 80}
024016 {\bf \ }

Olmo G J 2011 {\it Int. J. Mod. Phys}. D{\bf 20} 413 {\em \ }

\bibitem{} {\bf \ }Sotiriou T P and Liberati S 2007 {\it Annals Phys.} {\bf %
322} 935

Sotiriou T P and Liberati S 2007 {\it J. Phys. Conf. Ser.} {\bf 68} 012022

\bibitem{} Sotiriou T P 2009 {\it Class. Quant. Grav.} {\bf 26} 152001

\bibitem{} Vitagliano{\bf \ }V, Sotiriou T P and Liberati S 2010 {\it Phys.
Rev.} D{\bf 82} 084007

Koivisto{\bf \ }T S{\bf \ }2011 {\it Phys. Rev. }D{\bf 83} 101501

Olmo G J, Sanchis-Alepuz H and Tripathi S arXiv:1002.3920 [gr-qc].{\bf \ }

\bibitem{} Capozziello S, Cianci R, Stornaiolo C and Vignolo S 2007 {\it %
Class. Quant. Grav}. {\bf 24} 6417

\bibitem{} Sotiriou T P 2009 {\it Class. Quant. Grav}. {\bf 26} 152001

Vitagliano V, Sotiriou T P and Liberati S 2011{\em \ }{\it Annals Phys.}
{\bf 326} 1259 {\bf \ }

\bibitem{} Shapiro I 2002 {\it Phys. Rept.} {\bf 357} 113

\bibitem{} Hammond R 2002 {\it Rept. Prog. Phys}. {\bf 65} 599

\bibitem{} Hehl F W and Obukhov Y N arXiv:0711.1535 [gr-qc]

\bibitem{} Hammond R 2010 {\it Gen. Rel. Grav}. {\bf 42} 2345

\bibitem{} Jimenez J B and Koivisto T S arXiv:1201.4018 [gr-qc]

\bibitem{} Aldrovandi R, Pereira J G and Vu K H 2004 {\it Braz. J. Phys.}
{\bf 34} 1374 (arXiv:gr-qc/0312008)

\bibitem{} Bengochea G and Ferraro R 2009 {\it Phys. Rev.} D{\bf 79} 124019

Linder E V 2010 {\it Phys. Rev.} D{\bf 81} 127301

Li B, Sotiriou T P and Barrow J D 2011 {\it Phys. Rev}. D{\bf 83} 104017

\bibitem{} Harko T, Koivisto T S, Lobo F S N and Olmo G J arXiv:1110.1049
[gr-qc]

\bibitem{} Nojiri S and Odintsov S D 2003 {\it Phys. Rev.} D{\bf \ 68} 123512

M. Carroll M, Duvvuri V, Trodden M and Turner M S 2004 {\it Phys. Rev.} D%
{\bf \ 70} 043528

\bibitem{} Vollick D N 2003 {\it Phys. Rev.} D {\bf 68} 063510

\bibitem{} Barrow J D and Hervik S 2006 {\it Phys. Rev. }D{\bf 74} 124017

\bibitem{} Olmo G J arXiv:1112.1572 [gr-qc]

Barragan C and Olmo G J 2010 {\it Phys. Rev}. D{\bf 82} 084015

Olmo{\bf \ }G J, Sanchis-Alepuz H and Tripathi S 2009 {\it Phys. Rev}. D{\bf %
80} 024013

\bibitem{} Rubilar G F 1998 {\it Class. Quantum Grav}. {\bf 15} 239

\bibitem{} Hehl F W, McCrea J D, Mielke E W and Ne'emanY 1995 {\it Phys. Rep.%
} {\bf 258} 1

\bibitem{} Shie K-F, Nester J M and Yo H-J 2008 {\it Phys. Rev}. D {\bf 78}
023522

Chen H, Ho F-H, Nester J M, Wang C-H, and Yo H-J 2009 {\it JCAP} {\bf 0910}
027

Baekler P, Hehl F W and Nester J M 2011 {\it Phys. Rev.} D {\bf 83} 024001

Baekler P and Hehl F W arXiv:1105.3504 [gr-qc]

Li X-Z, Sun C-B and Xi P 2009 {\it Phys. Rev.} D {\bf 79} 027301

Ao X-C, Li X-Z and Xi P 2010 {\it Phys. Lett.} B{\bf 694} 186

Mielke E W and Romero E S 2006 {\it Phys. Rev.} D {\bf 73} 043521

\bibitem{} Garkun A S, Kudin V I, Minkevich AV and Vasilevsky Y G
arXiv:1107.1566 [gr-qc]

Minkevich A V arXiv:1102.0620 [gr-qc]

Minkevich A V 2011 {\it Mod. Phys. Lett.} A {\bf 26} 259

Minkevich A V 2009 {\it Phys. Lett.} B {\bf 678}, 423

Minkevich A V,Garkun A S and Kudin V I 2007 {\it Class. Quant. Grav.} {\bf 24%
} 5835

\bibitem{} Capozziello S, Martin-Moruno P and Rubano C 2010 {\it Phys. Lett}%
. B {\bf 689} 117

Faraoni V and Gunzig E 1999 {\em \ }{\it Int. J. Theor. Phys.} {\bf 38} 217

\bibitem{} Wainwright J and Ellis G F R 1997 {\it Dynamical System in
Cosmology,} (Cambridge University Press, Cambridge)

Coley A A, arXiv:gr-qc/9910074

Carloni S and Dunsby P K S arXiv:gr-qc/0611122

Carloni S, Troisi A and Dunsby P K S arXiv:0706.0452

Faraoni V 2005 {\it Annals Phys.} {\bf 317} 366

Faraoni V 2005 {\it Phys. Rev}. D {\bf 72} 061501

Faraoni V and Nadeau S 2005 {\it Phys. Rev.} D {\bf 72} 124005

\bibitem{} Wands D and Holden D J 1998 {\it Class. Quant. Grav}. {\bf 15}
3271

Copeland E J, Liddle A R and Wands D 1998 {\it Phys. Rev. }D {\bf 57} 4686

Coley A A 1999 {\it Gen. Relativ. Grav}. {\bf 31} 1295

Gunzig E et al 2000 {\it Class. Quant. Grav.} {\bf 17} 1783

\bibitem{} Tsamparlis M 1979 {\it Phys. Lett}. A {\bf 75} 27

Goenner H F M and Muller-Hoissen F 1984 {\it Class. Quant. Grav}. {\bf 1},
651

\bibitem{} Deser S and Tekin B 2002 {\it Phys. Rev. Lett}. {\bf 89} 101101

\bibitem{} Mannheim P D arXiv:1101.2186 [hep-th]

Maldacena J arXiv:1105.5632 [hep-th]

\bibitem{} Lu H and Pope C N 2011 {\it Phys. Rev. Lett}. {\bf 106}, 181302

Deser S, Liu H, Lu H, C N Pope, Sisman T C and B. Tekin 2011 {\it Phys. Rev}%
. D {\bf 83} 061502

Lu H, Pang Y and Pope C N arXiv:1106.4657 [hep-th]

Chen Y-X, Lu H and Shao K-N arXiv:1108.5184 [hep-th]



\end{thebibliography}
\end{document}